\documentclass{article} %
\usepackage{iclr2019_conference,times}

\iclrfinalcopy

\PassOptionsToPackage{numbers, compress}{natbib}

\usepackage{booktabs} %
\usepackage{lineno, blindtext, color}

\usepackage{tcolorbox}
\usepackage{environ}
\usepackage{xargs}                      %

\usepackage{listings}

\definecolor{oldeditorcolor}{rgb}{0.215686,0.494118,0.721569}

\newif\ifsubmit
\submittrue
\ifsubmit
    \usepackage[disable]{todonotes} %
    \newcommand{\usure}[1]{}
    \newcommand{\change}[1]{}
    \newcommand{\info}[1]{}
    \newcommand{\improvement}[1]{}
    \newcommandx{\cheng}[2][1=]{}
    \newcommandx{\abdul}[2][1=]{}
    \newcommandx{\abi}[2][1=]{}
    \newcommandx{\renee}[2][1=]{}
    \newcommandx{\junjun}[2][1=]{}
    \newcommandx{\wenmei}[2][1=]{}
    \newcommandx{\old}[2][1=]{}
    \newcommandx{\delete}[2][1=]{}
\else
\usepackage{comment}
\usepackage[
    colorinlistoftodos,
    prependcaption,
    draft,
    textsize=tiny,
    obeyFinal,
    textwidth=2cm
]{todonotes}
	 
	\newcounter{adtodocounter} 
	\newcounter{cltodocounter} 
	\newcounter{jjtodocounter} 
	\newcounter{wmhtodocounter} 
	\newcounter{abtodocounter} 
	\newcounter{rntodocounter} 
	\newcounter{oldtodocounter} 
	\newcounter{deletetodocounter} 
    \newcommandx{\unsure}[2][1=]{\todo[linecolor=red,backgroundcolor=red!25,bordercolor=red,#1]{#2}}
    \newcommandx{\change}[2][1=]{\todo[linecolor=blue,backgroundcolor=blue!25,bordercolor=blue,#1]{#2}}
    \newcommandx{\info}[2][1=]{\todo[linecolor=OliveGreen,backgroundcolor=OliveGreen!25,bordercolor=OliveGreen,#1]{#2}}
    \newcommandx{\improvement}[2][1=]{ \marginpar[\todo[linecolor=Plum,backgroundcolor=Plum!25,bordercolor=Plum,#1]{#2}]{}}
    \newcommandx{\thiswillnotshow}[2][1=]{\todo[disable,#1]{#2}}
    \newcommandx{\abi}[2][1=]{\stepcounter{abtodocounter} \todo[linecolor=red,backgroundcolor=purple!25,bordercolor=purple,#1]{AB(\thecltodocounter): #2}}
    \newcommandx{\abdul}[2][1=]{\stepcounter{adtodocounter} \todo[linecolor=red,backgroundcolor=red!25,bordercolor=red,#1]{AD(\theadtodocounter): #2}}
    \newcommandx{\cheng}[2][1=]{\stepcounter{cltodocounter} \todo[linecolor=teal,backgroundcolor=teal!25,bordercolor=teal,#1]{CL(\theabtodocounter): #2}}
    \newcommandx{\renee}[2][1=]{\stepcounter{rntodocounter} \todo[linecolor=yellow,backgroundcolor=yello!25,bordercolor=OliveGreen,#1]{RN(\therntodocounter): #2}}
    \newcommandx{\jinjun}[2][1=]{\stepcounter{jjtodocounter} \todo[linecolor=yellow,backgroundcolor=yellow!25,bordercolor=yellow,#1]{JJ(\thejjtodocounter): #2}}
    \newcommandx{\wenmei}[2][1=]{\stepcounter{wmhtodocounter} \todo[linecolor=Plum,backgroundcolor=Plum!25,bordercolor=Plum,#1]{WH(\thewmhtodocounter): #2}}
    \newcommandx{\old}[2][1=]{\stepcounter{oldtodocounter} \todo[linecolor=oldeditorcolor,backgroundcolor=oldeditorcolor!25,bordercolor=oldeditorcolor,#1]{OLD(\theoldtodocounter): #2}}
    \newcommandx{\delete}[2][1=]{\stepcounter{deletetodocounter} \todo[linecolor=oldeditorcolor,backgroundcolor=oldeditorcolor!25,bordercolor=oldeditorcolor,#1]{OLD(\thedeletetodocounter): #2}}
\fi

\definecolor{yamlred}{rgb}{0.843137,0.188235,0.152941}
\definecolor{yamlblue}{rgb}{0.27451,0.32549,0.384314}
\definecolor{yamlkey}{rgb}{0.980392,0.501961,0.447059}

\newcommand\YAMLfontsize{\fontsize{6}{6}}
\newcommand\YAMLfontstyle{\ttfamily}
\newcommand\YAMLcolonstyle{\color{black}\mdseries\YAMLfontsize\YAMLfontstyle}
\newcommand\YAMLkeystyle{\bfseries\color{yamlblue}\YAMLfontsize\YAMLfontstyle}
\newcommand\YAMLvaluestyle{\color{black}\mdseries\YAMLfontsize\YAMLfontstyle}

\definecolor{yamlgray}{rgb}{0.5,0.5,0.5}
\lstdefinelanguage{yaml}
{
  keywords={true,false,null,y,n},
  keywordstyle=\YAMLvaluestyle,
  basicstyle=\YAMLfontsize\YAMLkeystyle,                                 %
  sensitive=false,
  comment=[l]{\#},
  morecomment=[s]{/*}{*/},
  commentstyle=\YAMLcolonstyle\color{yamlred}\YAMLfontstyle\YAMLfontsize\YAMLfontstyle,
  stringstyle=\YAMLvaluestyle\YAMLfontstyle\YAMLfontsize\YAMLfontstyle,
  moredelim=[l][\color{orange}]{\&},
  moredelim=[l][\color{yamlred}]{*},
  moredelim=**[il][\YAMLcolonstyle{:}\YAMLvaluestyle]{:},   %
  morestring=[b]',
  morestring=[b]",
  literate =    {---}{{\ProcessThreeDashes}}3
                {>}{{\textcolor{yamlred}\textbar}}1     
                {|}{{\textcolor{yamlred}\textbar}}1 
                {\ -\ }{{\mdseries\ -\ }}3,       
    frame=top,
    frame=bottom,
    numbers=left,
    framerule=1pt,
    rulesep=1pt,
	breaklines=true,
    mathescape=true,
    xleftmargin=0em,
    framexleftmargin=0em,
    xrightmargin=0pt,
    stepnumber=1,
    escapechar=|,
    captionpos=t,
    escapeinside={(*}{*)},
    numberstyle=\ttfamily\tiny\color{yamlgray},
}

\makeatother

\newcommand\ProcessThreeDashes{\llap{\color{cyan}\mdseries-{-}-}}

\definecolor{pbgray}{rgb}{0.5,0.5,0.5}
\lstdefinelanguage[2]{protobuf}{%
    sensitive=true,%
    numbers=left,
	breaklines=true,
    mathescape=true,
    xleftmargin=2em,
    framexleftmargin=3em,
    stepnumber=1,
    escapechar=|,
    keywordstyle=\YAMLvaluestyle,
    basicstyle=\YAMLfontsize\YAMLkeystyle,  
    commentstyle=\YAMLcolonstyle\color{yamlred}\YAMLfontstyle\YAMLfontsize\YAMLfontstyle,
    stringstyle=\YAMLvaluestyle\YAMLfontstyle\YAMLfontsize\YAMLfontstyle,
    numberstyle=\ttfamily\tiny\color{yamlgray}, 
    frame=single,
    morecomment=[l]{//},%
    morestring=[b]{"},%
    morekeywords={enum,oneof,map,syntax,public,import,option,package,message,%
        group,optional,required,repeated,default,reserved,extend,extensions,%
        to,max,service,rpc,returns,true,false},%
    morekeywords=[2]{%
        double,float,int32,int64,uint32,uint64,sint32,sint64,%
        fixed32,fixed64,sfixed32,sfixed64,bool,string,bytes},%
    morekeywords=[3]{%
        deprecated, uninterpreted_option,%
        java_package,java_outer_classname,java_multiple_files,%
        java_generate_equals_and_hash,java_string_check_utf8,optimize_for,%
        go_package,cc_generic_services,java_generic_services,%
        py_generic_services,cc_enable_arenas,obj_class_prefix,%
        csharp_namespace,%
        message_set_wire_format,no_standard_descriptor_accessor,map_entry,%
        ctype, packed,jstype,lazy,weak,%
        allow_alias}%
}
\lstalias[]{protobuf2}[2]{protobuf}

\lstdefinelanguage[3]{protobuf}[2]{protobuf}{%
    frame=top,
    frame=bottom,
    numbers=left,
    framerule=1pt,
    rulesep=1pt,
	breaklines=true,
    mathescape=true,
    xleftmargin=0em,
    framexleftmargin=0em,
    xrightmargin=0pt,
    stepnumber=1,
    escapechar=|,
    captionpos=t,
    deletekeywords={
        group,%
        extensions, to, extend, max,%
        required,%
        optional,%
        default}%
}
\lstalias[]{protobuf3}[3]{protobuf}  %

\newcommand{\xparsep}{$-$}

\usepackage[hidelinks]{hyperref}
\hypersetup{draft}

\usepackage{latexsym}
\usepackage{verbatim}
\usepackage{graphicx}
\usepackage{subcaption} 
\usepackage[font={small}]{caption}
\usepackage{booktabs} %
\usepackage{soul}
\usepackage{amsthm}
\usepackage{amsmath}
\usepackage[utf8]{inputenc}
\usepackage{xspace}
\usepackage{array}
\usepackage{makecell}
\usepackage{rotating}
\usepackage{filecontents}
\usepackage{pgfplots}
\usepackage{pgfplotstable}
\usepackage{tikz}
\usepackage[normalem]{ulem}
\usepackage{enumitem}
\usepackage{amssymb}%

\usepackage{times}
\usepackage{url}
\makeatletter
\newsavebox{\measure@tikzpicture}
\NewEnviron{scaletikzpicturetowidth}[1]{%
  \def\tikz@width{#1}%
  \begin{lrbox}{\measure@tikzpicture}%
  \BODY
  \end{lrbox}%
  \pgfmathparse{#1/\wd\measure@tikzpicture}%
  \BODY
}

\setlist{noitemsep,nolistsep}

\newcommand{\cmmnt}[1]{\ignorespaces}

\definecolor{myred}{rgb}{0.843137,0.188235,0.152941}
\definecolor{myblack}{rgb}{0.27451,0.32549,0.384314}
\definecolor{mygreen}{rgb}{0.301961,0.686275,0.290196}
\definecolor{myyellow}{rgb}{0.996078,0.878431,0.564706}
\definecolor{myblue}{rgb}{0.568627,0.74902,0.858824}

\pgfplotsset{compat=newest,}
\usepgfplotslibrary{external}
\usetikzlibrary{babel}

\pgfplotsset{every axis/.style={scale only axis}}

\definecolor{plotcolor1}{rgb}{0.568627,0.74902,0.858824}
\definecolor{plotcolor2}{rgb}{0.996078,0.878431,0.564706}
\definecolor{plotcolor3}{rgb}{0.27451,0.32549,0.384314}
\definecolor{plotcolor4}{rgb}{0.843137,0.188235,0.152941}
\definecolor{plotcolor5}{rgb}{0.988235,0.552941,0.34902}
\definecolor{plotcolor6}{rgb}{0.596078,0.305882,0.639216}
\definecolor{plotcolor7}{rgb}{0.65098,0.337255,0.156863}
\definecolor{plotcolor8}{rgb}{0.105882,0.619608,0.466667}
\definecolor{plotcolor9}{rgb}{1.,1.,0.6}
\definecolor{plotcolor10}{rgb}{0.745098,0.729412,0.854902}
\definecolor{plotcolor11}{rgb}{0.984314,0.501961,0.447059}
\definecolor{plotcolor12}{rgb}{0.501961,0.694118,0.827451}
\definecolor{plotcolor13}{rgb}{1.,1.,0.2}
\definecolor{plotcolor14}{rgb}{0.992157,0.705882,0.384314}
\definecolor{plotcolor15}{rgb}{0.988235,0.803922,0.898039}
\definecolor{plotcolor16}{rgb}{0.701961,0.870588,0.411765}
\definecolor{plotcolor17}{rgb}{0.215686,0.494118,0.721569}
\definecolor{plotcolor18}{rgb}{0.941176,0.231373,0.12549}
\definecolor{plotcolor19}{rgb}{0.168627,0.54902,0.745098}
\definecolor{plotcolor20}{rgb}{0.552941,0.827451,0.780392}
\definecolor{plotcolor21}{rgb}{0.968627,0.505882,0.74902}
\definecolor{plotcolor22}{rgb}{0.6,0.6,0.6}
\definecolor{plotcolor23}{rgb}{0.301961,0.686275,0.290196}
\definecolor{plotcolor24}{rgb}{0.980392,0.501961,0.447059}

\usepgfplotslibrary{colorbrewer}
\pgfplotsset{cycle list/Dark2-8}

\newcommand{\carml}{MLModelScope\xspace}

\newcommand{\ignore}[1]{}

\DeclareRobustCommand*\circled[1]{\tikz[baseline=(char.base)]{
            \node[shape=circle,white,draw,fill=black,inner sep=0.0pt] (char) {\small #1};}}

\newcommand*\circledwhite[1]{\tikz[baseline=(char.base)]{
            \node[shape=circle,draw,inner sep=0.0pt] (char) {\small  #1};}}

\newcommand{\floor}[1]{\left\lfloor #1 \right\rfloor}

\newcolumntype{C}{>{\centering\arraybackslash} m{.1\linewidth} }  %

\usepackage{times}
\usepackage{epsfig}
\usepackage{graphicx}
\usepackage{amsmath}
\usepackage{amssymb}
\usepackage{enumitem}
\usepackage{enumitem}
\usepackage{kantlipsum}
\usepackage{adjustbox}

\usepackage[utf8]{inputenc} %
\usepackage[T1]{fontenc}    %
\usepackage{hyperref}       %
\usepackage{url}            %
\usepackage{booktabs}       %
\usepackage{amsfonts}       %
\usepackage{nicefrac}       %
\usepackage{microtype}      %

\usepackage{xspace}
\usepackage{filecontents}
\usepackage{pgfplots}
\usepackage{pgfplotstable}
\usepackage{multirow}
\usepackage{tikz}
\usepackage[normalem]{ulem}
\usepackage{enumitem}
\usepackage{listings}
\usepackage{tcolorbox}
\usepackage{xargs}                      %
\usepackage{amssymb}%
\usepackage{pifont}%

\usepackage{wrapfig}
\usepackage{caption}
\usepackage{subcaption}
\usepackage{environ}%

\tcbuselibrary{listings,breakable}

\definecolor{myred}{rgb}{0.843137,0.188235,0.152941}
\definecolor{myblack}{rgb}{0.27451,0.32549,0.384314}
\definecolor{mygreen}{rgb}{0.301961,0.686275,0.290196}
\definecolor{myyellow}{rgb}{0.996078,0.878431,0.564706}
\definecolor{myblue}{rgb}{0.568627,0.74902,0.858824}

\input{glyphtounicode}
\DeclareRobustCommand*\level[1]{\tikz[baseline=(char.base)]{
            \node[rounded corners,white,draw,fill=black,line width=1pt,rounded corners=2pt,inner sep=1.0pt] (char) {\normalfont\sffamily\textsf{L#1}};}}

\DeclareRobustCommand*\circled[1]{\tikz[baseline=(char.base)]{
            \node[circle,white,,draw,fill=black,inner sep=0.0pt] (char) {\normalfont\footnotesize\sffamily\textsf{#1}};}}
\DeclareRobustCommand*\circledwhite[1]{\tikz[baseline=(char.base)]{
            \node[circle,draw,inner sep=0.0pt] (char) {\normalfont\footnotesize\sffamily\textsf{#1}};}}

\newcommand{\feature}\ignore

\title{MLModelScope: A Distributed Platform for Model Evaluation and Benchmarking at Scale}

\begin{document}

\author{Abdul Dakkak\textsuperscript{*},Cheng Li\textsuperscript{*} \\
University of Illinois, Urbana-Champaign \\
\texttt{\{dakkak,cli99\}@illinois.edu}
\And
Jinjun Xiong \\
IBM Thomas J. Watson Research Center \\
\texttt{jinjun@us.ibm.com}
\And
Wen-mei Hwu \\
University of Illinois, Urbana-Champaign \\
\texttt{w-hwu@illinois.edu}
}

\renewcommand{\thefootnote}{\fnsymbol{footnote}}
\footnotetext[1]{The two authors contributed equally to this paper.}

\date{}

\maketitle

\begin{abstract}

Machine Learning (ML) and Deep Learning (DL) innovations are being introduced at such a rapid pace that researchers are hard-pressed to analyze and study them.
The complicated procedures for evaluating innovations, along with the lack of standard and efficient ways of specifying and provisioning ML/DL evaluation, is a major ``pain point” for the community.
This paper proposes \carml, an open-source, framework/hardware agnostic, extensible and customizable design that enables repeatable, fair, and scalable model evaluation and benchmarking. 
We implement the distributed design with support for all major frameworks and hardware, and equip it with web, command-line, and library interfaces.
To demonstrate \carml's capabilities we perform parallel evaluation and show how subtle changes to model evaluation pipeline affects the accuracy and HW/SW stack choices affect performance.

\end{abstract}

\section{Introduction}\label{sec:intro}

The emergence of Machine Learning (ML) and Deep Learning (DL) within a wide array of application domains has ushered in a great deal of innovation in the form of new models and hardware/software (HW/SW) stacks (frameworks, libraries, compilers, and hardware accelerators) to support these models.
Being able to evaluate and compare these innovations in a timely manner is critical for their adoption.
These innovations are introduced at such a rapid pace~\citep{dean2018new, arXiv} that researchers are hard-pressed to study and compare them.
As a result, there is an urging need by both research and industry for a \textit{scalable} model/HW/SW evaluation platform.%

Evaluation platforms must maintain \textit{repeatability} (the ability to reproduce a claim) and \textit{fairness} (the ability to keep all variables constant and allow one to quantify and isolate the benefits of the target of interest).
For ML/DL, repeatable and fair evaluation is challenging, since there is a tight coupling between model execution and the underlying HW/SW components.
Model evaluation is a complex process where the model, dataset, evaluation method, and HW/SW stack must work in unison to maintain the accuracy and performance claims (e.g. latency, throughput, memory usage).
To maintain repeatability, authors are encouraged to publish their code, containers, and write documentation which details the usage along with HW/SW requirements~\citep{mitchell2019model,ReproducibilityChecklist,showyourwork,lipton2019research,pineau2018iclr}.
Often, the documentation miss details which make the results not reproducible.
To perform a fair evaluation, evaluators have to manually normalize the underlying stack and delineate the codes to characterize performance or accuracy.
This is a daunting endeavor.
As a consequence, repeatable and fair evaluation is a ``pain-point'' within the community~\citep{gundersen2018reproducible,plesser2018reproducibility,ghanta2018systems, hutson2018artificial,li2019random,tatman2018practical,rml,icrl}.
Thus, an evaluation platform design must have a standard way to specify, provision, and introspect evaluations to guarantee repeatability and fairness.

In this paper, we propose \carml: a distributed design which consists of a specification and a runtime that enables repeatable, fair, and scalable evaluation and benchmarking.
The proposed specification is text-based and encapsulates the model evaluation by defining its pre-processing, inference, post-processing pipeline steps and required SW stack.
The runtime system uses the evaluation specification along with user-defined HW constraints as input to provision the evaluation, perform benchmarking, and generate reports.
More specifically, \carml{} guarantees \feature{1} repeatable and \feature{2} fair evaluation by
(1) defining a novel scheme to specify model evaluation which separates the entanglement of data/code/SW/HW;
(2) defining common techniques to provision workflows with specified HW/SW stacks; and (3) providing a consistent benchmarking and reporting methodology.
Through careful design, \carml solves the design objectives while being framework/hardware agnostic, extensible, and customizable.

In summary, this paper makes the following contributions:
\circledwhite{1} we comprehensively discuss the complexity of model evaluation and describe prerequisites for a model evaluation platform.
\circledwhite{2} We propose a model evaluation specification and an open-source, framework/hardware agnostic, extensible, and customizable distributed runtime design which consumes the specification to execute model evaluation and benchmarking at scale.
\circledwhite{3} We implemented the design with support for
Caffe, Caffe2, CNTK, MXNet, PyTorch, TensorFlow, TensorRT, and TFLite, running on ARM, Power, and x86 with CPU, GPU, and FPGA.
\circledwhite{4} For ease of use, we equip \carml with command line, library, and ready-made web interfaces which allows ``push-button'' model evaluation\footnote{A video demo of web UI is at https://drive.google.com/open?id=1LOXZ7hs\_cy-i0-DVU-5FfHwdCd-1c53z.}.
\circledwhite{5} We also add introspection capability in \carml to analyze accuracy at different stages and capture latency and memory information at different levels of the HW/SW stack.
\circledwhite{6} We showcase \carml{} by running experiments which compare different model pipelines, hardware, and frameworks.

\vspace{-3pt}

\begin{wrapfigure}{ht}{0.4\textwidth}
  \vspace{-55pt}
  \begin{center}
    \setlength{\abovecaptionskip}{5pt}
  \includegraphics[clip, width=0.4\textwidth]{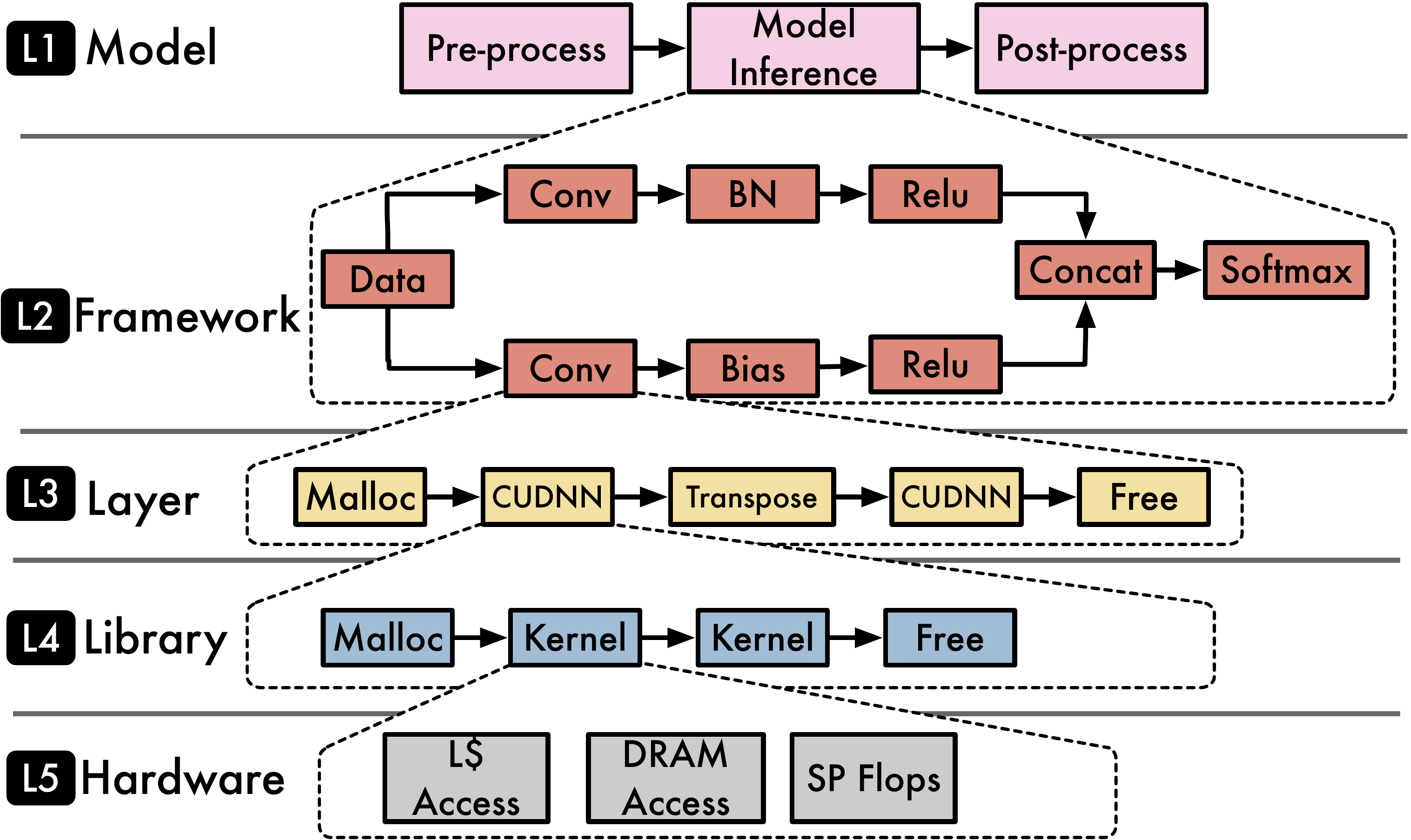}
  \caption{Execution of a model evaluation at different levels of hardware and software abstractions on GPUs.
  }
  \label{fig:profile_levels}
  \end{center}
  \vspace{-10pt}
\end{wrapfigure}

\section{Model Evaluation Challenges}\label{sec:motivation}

Model evaluation is complex.
Researchers that publish and share DL models can attest to that but are sometimes unaware of the full scope of this complexity.
To perform repeatable and fair evaluation, we need to be cognizant of the HW/SW stack and how it affects the accuracy and performance of a model.
Figure~\ref{fig:profile_levels} shows our classification of the HW/SW stack levels.
\textbf{Model level} (\level{1}) evaluates a model by performing input pre-processing, model inference, and post-processing.
The pre-processing stage transforms the user input into a form that the model expects.
The model inference stage calls the framework's inference API on the processed input and produces an output.
The post-processing stage transforms the model output to a form that can be viewed by a user or used to compute metrics.
\textbf{Framework level} (\level{2}) performs model inference by executing the layers in the model graph using a framework such as TensorFlow, MXNet, or PyTorch.
\textbf{Layer level} (\level{3}) executes a sequence of ML library calls for layers such as convolution, normalization, or softmax.
\textbf{ML Library level} (\level{4}) invokes a chain of system library calls for functions in ML libraries such as cuDNN\citep{cudnn}, MKL-DNN~\citep{mkldnn} or OpenBLAS~\citep{xianyi2014openblas}.
And, last but not the least, at the \textbf{hardware level} (\level{5}), there are CPU/GPU instructions, disk, and network I/O events, and other low-level system operations through the entire model evaluation.
All the HW/SW abstractions must work in unison to maintain the reported accuracy and performance claims.
When things go awry, each level within the abstraction hierarchy can be suspect. %

Currently, model authors distribute models by publishing documentation and  ad hoc scripts to public repositories such as GitHub.
Due to the lack of specification, authors may under-specify or omit key aspects of  model evaluation.
This inhibits, or makes it difficult, for others to repeat their evaluations or validate their claims.
Thus all aspects of the model evaluation must be captured by a evaluation platform to guarantee repeatability.
To highlight this, consider the model evaluation pipeline at \level{1}.
While the model inference stage is relatively straight forward, the pre- and post-processing stages are surprisingly subtle and can easily introduce discrepancies in the results.
Some of the discrepancies might be ``silent errors'' --- where the evaluation is correct for the majority of the inputs but is incorrect for a small number of cases.
In general, accuracy errors due to under-specifying pre- and post-processing are difficult to identify and even more difficult to debug.
In Section~\ref{sec:pitfalls}, we show the effects of under-specifying different operations in pre-processing on image classification models.

The current practice of publishing models also causes a few challenges which must be addressed by a fair and scalable evaluation platform.
First, any two ad hoc scripts do not adhere to a consistent evaluation API.
The lack of a consistent API makes it difficult to evaluate models in parallel and, in turn, slows down the ability to quickly compare models across different HW/SW stacks.
Second, ad hoc scripts tend to not clearly demarcate the stages of the model evaluation pipeline.
This makes it hard to introspect and debug the evaluation.
Furthermore, since an apple-to-apple comparison between models requires a fixed HW/SW stack, it is difficult to perform honest comparison between two shared models without modifying some ad hoc scripts.
\carml{} addresses these challenges through the careful design of a model evaluation specification and a distributed runtime as described in Section~\ref{sec:design}.

\vspace{-3pt}

\begin{wrapfigure}{r}{0.42\textwidth}
     \vspace{-60pt}
    \centering%
\begin{adjustbox}{max width=0.38\textwidth}
\begin{lstlisting}[
    language=yaml,
    escapeinside={(*}{*)},
    floatplacement=H,
    escapechar=|,
    xleftmargin=0pt,
    xrightmargin=0pt,
    framerule=1pt,
    numbersep=4pt,
    tabsize=2,
]
|\label{line:start_model_id}|name: Inception-v3 # model name
version: 1.0.0 # semantic version of model 
|\label{line:end_model_id}|task: classification # model modality
license: MIT # model license
description: ...
|\label{line:start_framework_id}|framework: # framework information
  name: TensorFlow
|\label{line:end_framework_id}|  version: ^1.x # framework version constraint
|\label{line:start_container}|container: # containers used for architecture
  arm64: mlms/tensorflow:1-13-0_arm64-cpu
  amd64:
    cpu: mlms/tensorflow:1-13-0_amd64-cpu
    gpu: mlms/tensorflow:1-13-0_amd64-gpu
  ppc64le:
    cpu: mlms/tensorflow:1-13-0_ppc64le-cpu
|\label{line:end_container}|    gpu: mlms/tensorflow:1-13-0_ppc64le-gpu
|\label{line:start_env}|envvars:
|\label{line:end_env}|  - TF_ENABLE_WINOGRAD_NONFUSED: 0
|\label{line:start_inputs}|inputs: # model inputs
  - type: image  # first input modality
    layer_name: data
|\label{line:end_inputs}|    element_type: float32
|\label{line:start_pre_processing}|pre-processing: >
  def pre_processing(env, inputs):
    ... #e.g. import opencv as cv 
|\label{line:end_pre_processing}|    return preproc_inputs
|\label{line:start_outputs}|outputs: # model outputs
  - type: probability # output modality
    layer_name: prob
|\label{line:end_outputs}|    element_type: float32
|\label{line:start_post_processing}|post-processing: >
  def post_processing(env, inputs):
    ... # e.g. os.exec("Rscript ~/postproc.r")
|\label{line:end_post_processing}|    return postproc_inputs
|\label{line:start_weights}|source: # model source
|\label{line:end_weights}|  graph_path: https://.../inception_v3.pb
|\label{line:start_attrs}|training_dataset:  # dataset used for training
  name: ILSVRC 2012
|\label{line:end_attrs}|  version: 1.0.0 
\end{lstlisting}
\end{adjustbox}
  \vspace{-.1in}
  \captionof{lstlisting}{Example evaluation manifest.
  }
  \vspace{-40pt}
  \label{lst:model_manifest}
  \end{wrapfigure}

\section{\carml{} Design}\label{sec:design}

We propose \carml{}, an open-source, framework/hardware agnostic, extensible and customizable distributed system design to perform model evaluation and benchmarking at scale.
\carml consists of a model evaluation specification and a distributed runtime.

\subsection{Model Evaluation Manifest}\label{sec:manifest}

All models in \carml{} are described using a model specification (called \textit{manifest}).
The manifest is a text file and includes the information needed to run a model.
The manifest specifies information such as the model pre- and post-processing steps, its model sources (graph and weight), and its software stack. %
The hardware details are not present in the manifest, but are user-provided options when performing the evaluation. %
By decoupling the hardware specification from the manifest, a manifest can work across hardware.

An example manifest is shown in Listing~\ref{lst:model_manifest} and contains model name, version, and type of task (Lines~\ref{line:start_model_id}--\ref{line:end_model_id}); framework name and version constraints (Lines~\ref{line:start_framework_id}--\ref{line:end_framework_id}); containers to use for evaluation (Lines~\ref{line:start_container}--\ref{line:end_container}); model inputs (Lines~\ref{line:start_inputs}--\ref{line:end_inputs});  pre-processing function (Lines~\ref{line:start_pre_processing}--\ref{line:end_pre_processing}); model outputs (Lines~\ref{line:start_outputs}--\ref{line:end_outputs});  post-processing function (Lines~\ref{line:start_post_processing}--\ref{line:end_post_processing}); model resources (Lines~\ref{line:start_weights}--\ref{line:end_weights}); and other metadata attributes (Lines~\ref{line:start_attrs}--\ref{line:end_attrs}).
The key components of the manifest are:

\textbf{Software Stack}\xparsep \carml{} uses docker containers to maintain the software stacks. %
\carml{} provides ready-made containers for all popular frameworks, but users can use any container hosted on Docker Hub.
Multiple containers can be specified within the manifest.
The container used for evaluation is dependent on the executing hardware and whether the system has a GPU or not.

\textbf{Model Source}\xparsep Model source contains links to the model graph (the \verb|graph_path| field) and weights (the \verb|weights_path| field).
For frameworks which have one file to represent the graph and its weights, the weights field is omitted from the manifest.
The model can be stored in the cloud, downloaded on demand, and is cached to the local file system.

\begin{wrapfigure}{r}{0.3\textwidth}
  \vspace{-25pt}
  \centering%
\begin{adjustbox}{max width=0.25\textwidth}
\begin{lstlisting}[
    language=yaml,
    escapeinside={(*}{*)},
    floatplacement=H,
    escapechar=|,
    xleftmargin=0pt,
    xrightmargin=0pt,
    framerule=1pt,
    numbersep=4pt,
    tabsize=2,
]
type: image  # input modality
layer_name: data
element_type: float32
steps: # pre-processing steps
  decode:
    element_type: int8
    data_layout: NHWC
    color_layout: RGB
  crop:
    method: center
    percentage: 87.5
  resize:
    dimensions: [3, 299, 299]
    method: bilinear
    keep_aspect_ratio: true
  mean: [127.5, 127.5, 127.5]
  rescale: 127.5
\end{lstlisting}
\end{adjustbox}
\vspace{-.1in}
\captionof{lstlisting}{\carml's evaluation manifest for Inception-v3.
}
\vspace{-15pt}
\label{lst:image_model_manifest}
\end{wrapfigure}

\textbf{Versioning}\xparsep 
Models, frameworks, and datasets are all versioned within \carml{} using a semantic versioning~\citep{semanticver} scheme.
The versioning of frameworks and datasets supports constraints, such as \verb|^1.x| (Listing~\ref{lst:model_manifest},  Line~\ref{line:end_framework_id}).
This tells \carml{} that the model works on any TensorFlow v$1$ framework.

\textbf{Pre-/Post-Processing Functions}\xparsep To perform input pre-processing and output post-processing,  \carml{} allows arbitrary Python functions to be placed within the manifest file.
The pre- and post-processing functions have the signature \verb|def fun(env, data)| where \verb|env| contains metadata of the evaluation request and \verb|data| is a \verb|PyObject| representation of the user request for pre-processing and the model's output for post-processing.
Internally \carml{} executes the Python code within a Python sub-interpreter~\citep{python-subinterpreter} in the launched container.
To reduce data copy overhead parameters are passed by reference to the processing functions.
The pre- and post-processing functions are flexible; i.e. users may import external Python modules or invoke external scripts.
By allowing arbitrary pre- and post-processing function executions, \carml{} works with existing processing codes and is capable of supporting arbitrary modalities.

\textbf{Built-in Pre-/Post-Processing Functions}\xparsep 
As vision models are widely used and their pre- and post-processing operations are less diverse, \carml{} allows for common pre-processing image operations (e.g.\ image decoding, resizing, and normalization) and post-processing operations (e.g. topK, IOU, mAP, etc.) to be used within the manifest without writing code.
Internally, \carml{} invokes built-in pre- and post-processing code to perform these operations.
Listing~\ref{lst:image_model_manifest} can be placed within the inputs block (Lines~\ref{line:start_inputs}--\ref{line:end_inputs} in Listing~\ref{lst:model_manifest}) as the pre-processing steps for Inception-v3.
The steps are executed in the order that is specified, since, as we show in Section~\ref{sec:exper}, the order of operations can have a significant impact on achieved model accuracy.
Users are not required to use this feature, but using this feature allows users to easily compare pre- or post-processing steps. 
We use this mechanism during our evaluation in Section~\ref{sec:exper}.

\begin{wrapfigure}{r}{0.54\textwidth}
  \vspace{-55pt}
  \centering%
\begin{adjustbox}{max width=0.48\textwidth}
\begin{lstlisting}[
    language=protobuf2,
    escapeinside={(*}{*)},
    escapechar=|,
    xleftmargin=0pt,
    xrightmargin=0pt,
    framerule=1pt,
    numbersep=4pt,
    tabsize=2,
    frame=top,
    frame=bottom,
    numbers=left,
    framerule=1pt,
    rulesep=1pt,
	breaklines=true,
    mathescape=true,
    xleftmargin=0em,
    framexleftmargin=0em,
    xrightmargin=0pt,
    stepnumber=1,
    escapechar=|,
    captionpos=t,
    floatplacement=htbp
]
// Opens a predictor.
rpc ModelLoad(OpenRequest) returns (ModelHandle){}
// Close an open predictor.
rpc ModelUnload(ModelHandle) returns (CloseResponse){}
// Perform model inference on user data.
rpc Predict(PredictRequest) returns (PredictionResponse){}
\end{lstlisting}
\end{adjustbox}
\vspace{-.1in}
\captionof{lstlisting}{\carml's predictor RPC API consists of $3$ functions which are specified using Protobuf.}
\label{lst:prediction_api}
\vspace{-15pt}
\end{wrapfigure}

\subsection{The \carml{} Runtime}

The \carml{} runtime consumes the model manifest to provision evaluations and perform benchmarking.
Users evaluate a model by specifying its name, version, and framework along with the target hardware requirements.
The \carml{} runtime uses these user-provided constraints to query previous evaluations or schedule new ones.
The runtime is distributed and is built as a set of extensible and customizable modular components (see Figure~\ref{fig:arch}).
Due to space limitations, we only highlight the key components of the runtime (See Appendix for a description of all components):

\textbf{Framework Predictors}\xparsep
At the core of the software stack are the frameworks.
To enable uniform evaluation and maximize code reuse, \carml{} wraps each framework's C++ inference API to provide a uniform  interface (called \textit{predictor API}).
The predictor API (shown in Listing~\ref{lst:prediction_api}) is minimal and performs model loading, unloading, and inference.
So long as a program implements \carml{}'s predictor API, it can be plugged into the system.
This means that \carml{}'s design allows for exotic hardware or framework support.
For example, some hardware, such as FPGAs and ASICs, do not have a framework per se.
These hardware are exposed to \carml through a program which implements the predictor API.
The \texttt{ModelLoad} API for FPGAs, for example, downloads a bitfile and loads it onto the device.

The predictor API is linked against common code to perform container launching, manifest file handling, downloading of required assets, pre- and post-processing function execution, collecting of performance profiles, and publishing of results --- we call this bundle an \textit{agent}.
These agents can be run on separate machines, can be run in parallel, and are managed by the \carml{} orchestration layer.
Agents can be run on remote systems behind firewalls to allow for model evaluation on remote hardware --- this allows hardware providers to give model evaluators access to perform model evaluations without full unrestricted access to the hardware.
\carml{} does not require modifications to a framework and thus pre-compiled binary versions of frameworks (e.g. distributed through Python's pip) or customized versions of a framework work within \carml{}.

\textbf{Manifest and Predictor Registry}\xparsep
\carml{} uses a distributed key-value registry~\citep{escriva2012hyperdex} to store the model manifests and running agent information.
\carml{}'s orchestration layer leverages the registry to facilitate the discovery of models and  routing of user requests across the distributed agents using the HW/SW constraints provided by the user.
The registry is dynamic --- i.e. both model manifests and agents can be added and removed at runtime. %

\textbf{Profilers and Tracers}\xparsep
To enable performance debugging, \carml{} collects system, framework, and model level profiling information.
This data is published into a tracing server~\citep{opentracing,sigelman2010dapper} where it gets aggregated and summarized.
Through the trace, users get a holistic view of the performance of model evaluation and can identify bottlenecks.
To minimize overhead, the profilers are only active when a user enables them as part of the evaluation request.

\textbf{Web UI and Command Line Interface}\xparsep
Users interact with \carml{} through its web UI or command-line interface by specifying model and hardware constraints.
These constraints are used to query the database for previous evaluations or to schedule new ones.
Users can integrate \carml{} within their existing tools or pipelines by using its REST or RPC APIs. %

\begin{wrapfigure}{tr}{0.4\textwidth}
  \vspace{-35pt}
  \centering%
    \setlength{\abovecaptionskip}{5pt}
  \includegraphics[width=0.38\textwidth]{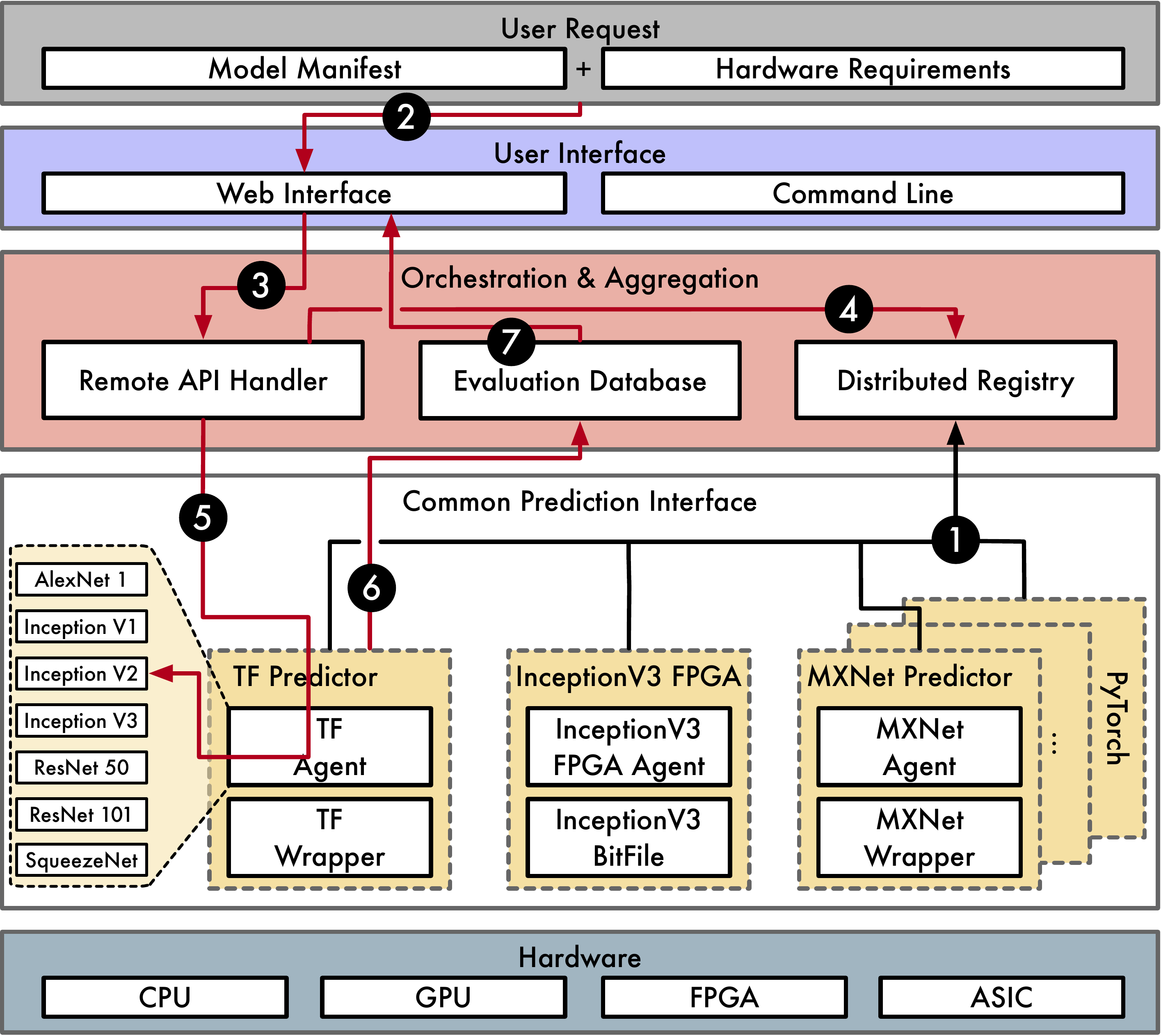}
  \caption{
  \carml's distributed runtime enables scalable  evaluation across models, frameworks, and systems.
  }
  \label{fig:arch}
 \vspace{-15pt}
\end{wrapfigure}

\subsection{\carml{} Evaluation Flow}\label{sec:flow}

To illustrate the execution flow of a model evaluation, consider a user wanting to run Inception-v3 trained using ILSVRC 2012 on an Intel system with TensorFlow satisfying the \texttt{"\(\geq\)1.10.x and \(\leq\)1.13.0"} version constraint.
The user specifies these constraints using \carml's UI and invokes the model evaluation. %
\carml{} then finds one or more systems which satisfy the user's constraints, sets up the environment, and launches the model evaluation within a container.
The results are then published to a database for subsequent analysis.

Figure~\ref{fig:arch} shows the evaluation flow of a user's request.
\circled{1}  On system startup, each agent publishes the hardware it is running on to the registry.
This information is made visible to the \carml{} orchestration layer.
\circled{2} A user then uses \carml's UI to request an evaluation by specifying the model, framework, and hardware constraints.
\circled{3} An API request is then performed to the remote API handler, which then \circled{4} queries the registry to find an agent which satisfies the user's constraints.
\circled{5} The request is then forwarded to one (or all) of the agents capable of running the evaluation.
The agents then provision the hardware and software environment and run the model.
\circled{6} The agents then collect and publish the results to a centralized evaluation database. %
\circled{7} Finally, an evaluation summary is presented to the user.

\vspace{-3pt}

\begin{figure}[ht]
  \centering
\begin{minipage}[t]{0.3\linewidth}
  \centering
    \setlength{\abovecaptionskip}{0pt}
  \includegraphics[width=.8\textwidth]{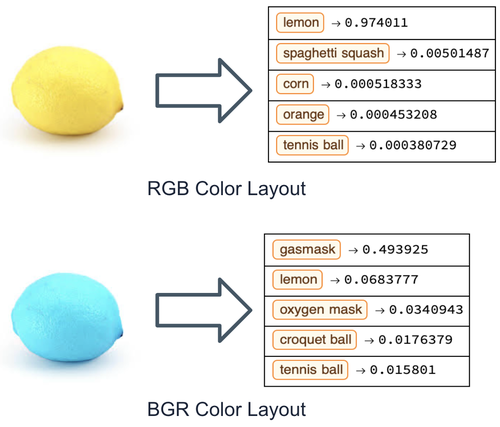}
  \caption{Top 5 predictions using Inception-v3 with RGB or BGR color layout.}
  \label{fig:lemon}
\end{minipage}%
  \hfill%
\begin{minipage}[t]{0.3\linewidth}
    \setlength{\abovecaptionskip}{0pt}
\centering
\includegraphics[width=1.0\textwidth]{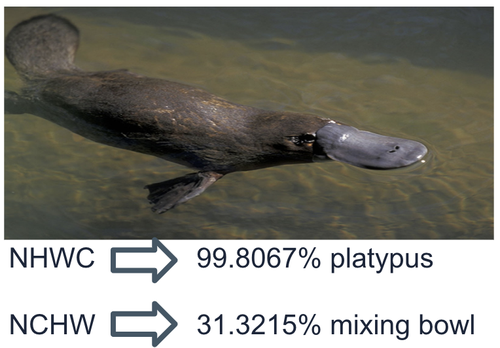}
\caption{Top 1 predictions using Inception-v3 with \texttt{NCHW} or \texttt{NHWC} data layout.}
\label{fig:platapus}
\end{minipage} %
  \hfill%
\begin{minipage}[t]{0.35\linewidth}
\centering
\includegraphics[width=1.0\textwidth]{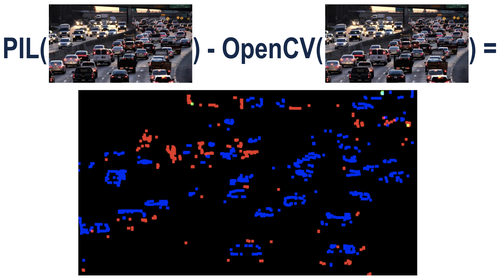}
\caption{Image decoding difference between PIL and OpenCV.}
\label{fig:pil_vs_opencv_diff}
\end{minipage} 

\begin{minipage}[t]{\linewidth}
\centering
    \setlength{\abovecaptionskip}{0pt}
\includegraphics[width=0.95\textwidth]{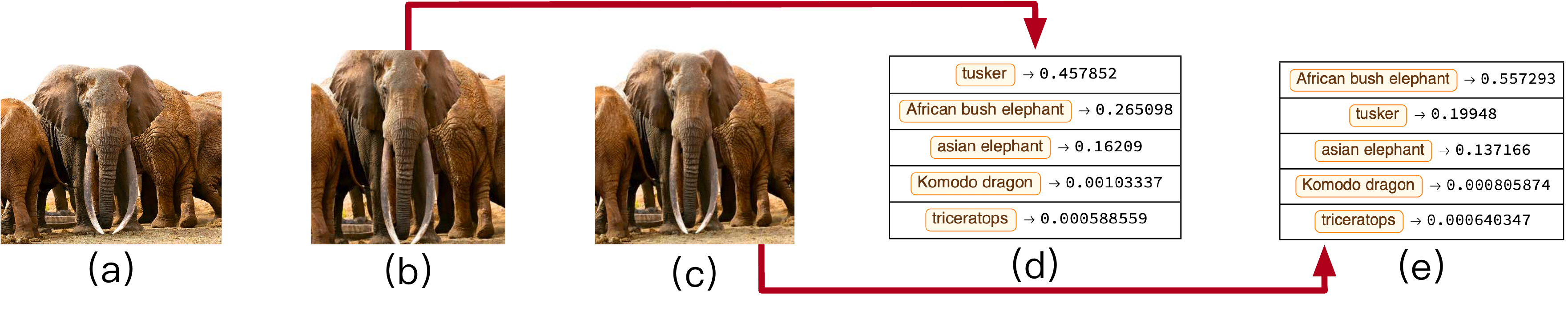}
\caption{Differences in the prediction results due to cropping using TensorFlow Inception-v3.}
\label{fig:diff_output_process_pred}
\end{minipage}

\begin{minipage}[t]{\linewidth}
\centering
    \setlength{\abovecaptionskip}{0pt}
\includegraphics[width=0.95\textwidth]{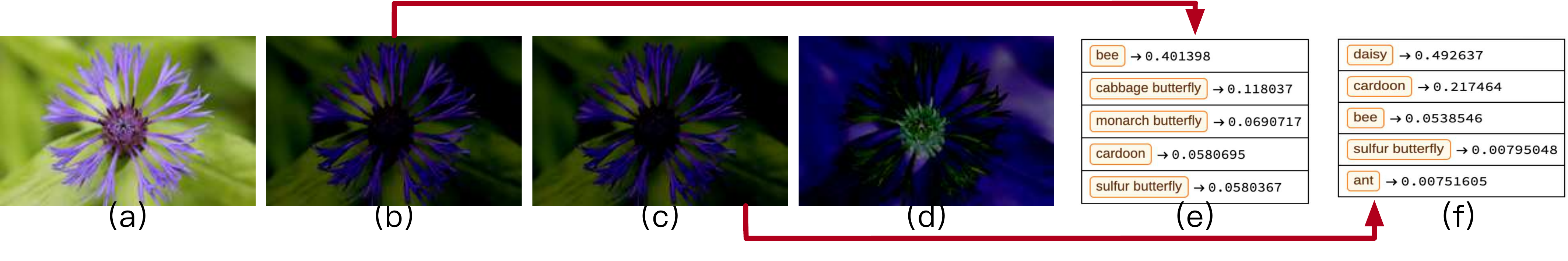}
\vspace{-5pt}
\caption{Differences due to order of operations using TensorFlow Inception-v3.}
\label{fig:diff_output_process}
\end{minipage}

\begin{minipage}[t]{\linewidth}
\centering
\vspace{-5pt}
    \setlength{\abovecaptionskip}{0pt}
\includegraphics[width=\textwidth]{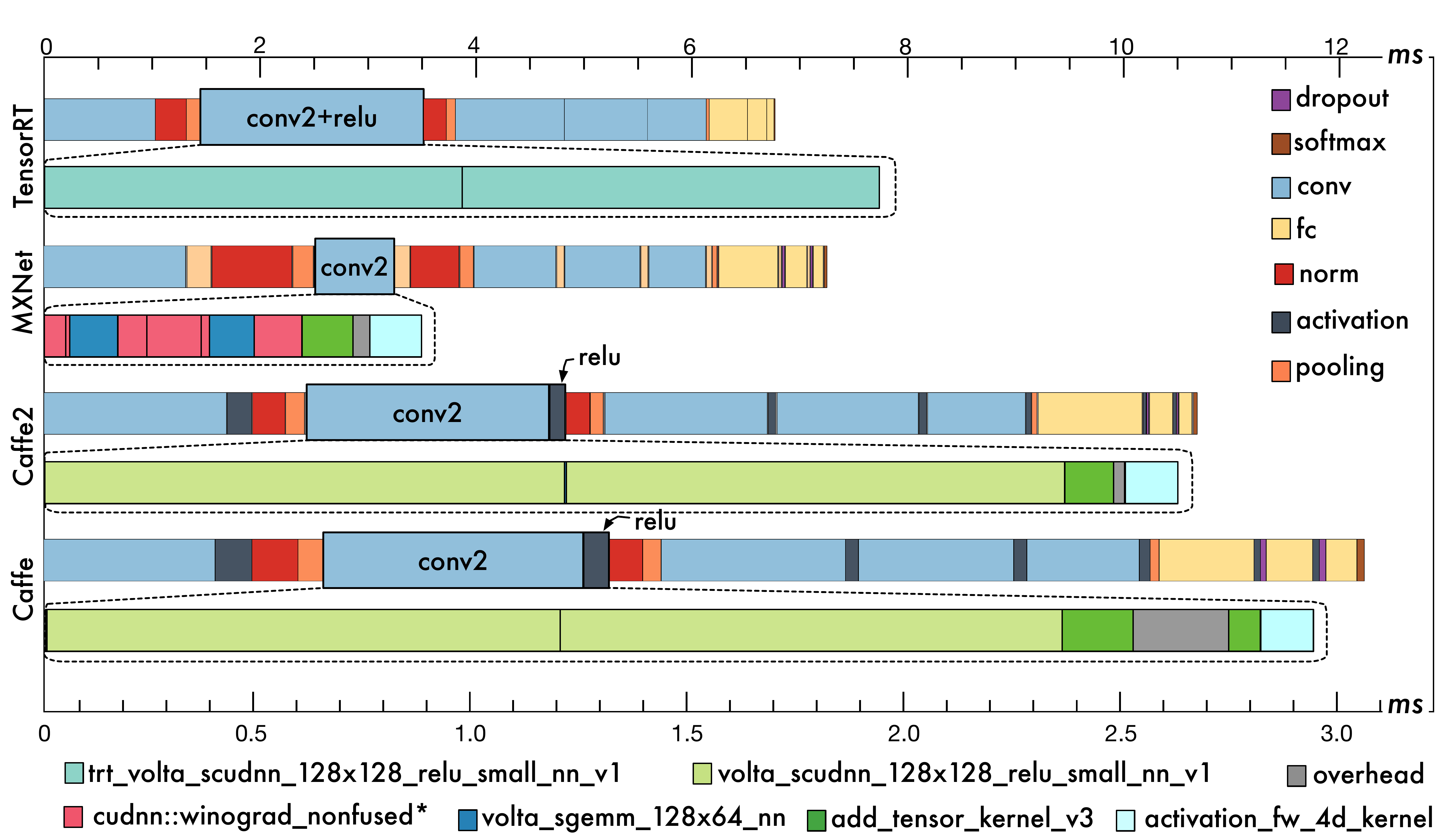}
\caption{
Performance of AlexNet with batch size $64$ across frameworks on Amazon P3. \carml{} enables one to understand and debug performance bottlenecks at layer and sub-layer granularity. 
The axis on top ($0–35$) is the duration ($ms$) to evaluate each layer within the model, while the axis at the bottom ($0–3$) is the duration ($ms$) to evaluate the kernels within the second convolution and Relu layers.
}
\label{fig:tensorrt_vs_caffe2_framework_compare}
\end{minipage}
\vspace{-20pt}

\end{figure}

\vspace{-3pt}

\section{Evaluation}\label{sec:exper}

We implemented the \carml{} design as presented in Section~\ref{sec:design} with support for popular frameworks (Caffe, Caffe2, CNTK, MXNet, PyTorch, TensorFlow, TensorRT, and TFLite) and tested it on common hardware (X86, PowerPC, and ARM CPUs as well as GPU and FPGA accelerators).
We populated it with over $300$ models covering a wide array of inference tasks such as image classification, object detection, segmentation, image enhancement, recommendation, etc. %
We considered three aspects of \carml{} for our evaluation: the effects of under-specified pre-processing on model accuracy, model performance across systems, and the ability to introspect model evaluation to identify performance bottlenecks.
To demonstrate \carml{}'s functionality, we installed it on multiple Amazon instances and performed the evaluation in parallel using highly cited image classification models.

Unless otherwise noted, all results use TensorFlow \texttt{1.13.0-rc2} compiled from source; CUDNN $7.4$; GCC $6.4.0$; Intel Core i7-7820X CPU with Ubuntu $18.04.1$; NVIDIA TITAN V GPU with CUDA Driver $410.72$; and CUDA Runtime $10.0.1$ (Amazon \texttt{p3.2xlarge} Instance).

\subsection{Model Pre-Processing}\label{sec:pitfalls}

We use \carml{} to compare models with different operations in the pre-processing stage.
Specifically, we look at the impact of image decoding, cropping, resizing, normalization, and data type conversion on model accuracy.
For all the experiments, the post-processing is a common operation which sorts the model's output %
to get the top \(K\) predictions.
To perform the experiments, we create variants of the original Inception-v3~\citep{silberman2016tensorflow,szegedy2016rethinking} pre-processing specification (shown in Listing~\ref{lst:image_model_manifest}).
We maintain everything else as constant with the exception to the operation of interest and evaluate the manifests through \carml's web UI. %

\textbf{Color Layout}\xparsep 
Models are trained with decoded images that are   in either RGB or BGR  layout.
For legacy reasons, \citet{opencv} decodes images in BGR layout by default and, subsequently, both Caffe and Caffe2 use the BGR layout~\citep{caffebgr}.
Other frameworks (such as TensorFlow and PyTorch) use RGB layout.
Intuitively, incorrect color layout only misclassifies images which are defined by their colors.
Images which are not defined by their colors, however, would be correctly classified.
Figure~\ref{fig:lemon} shows the Top 5 classifications for the same image when changing the color layout.

\textbf{Data Layout}\xparsep Images are represented by: $N$ (batch size), $C$ (channels), $H$ (height), and $W$ (width).
Models are trained using data in either \texttt{NCHW} or \texttt{NHWC} form.
Figure~\ref{fig:platapus} shows Inception-v3's (trained using \texttt{NHWC} layout)  Top1 result using different layouts for the same image.

\textbf{Decoding and Color Conversion}\xparsep
It is common to use JPEG as the image data serialization format (with ImageNet being stored as JPEG images).
Model developers use library functions such as \texttt{opencv.imread}, \texttt{PIL.Image.open}, or \texttt{tf.io.decode\_jpeg} to decode JPEG images.
These functions may use different decoding algorithms and color conversion methods. %
For example, we find the YCrCb to RGB color conversion to not be consistent across the PIL and OpenCV libraries.
Figure~\ref{fig:pil_vs_opencv_diff} shows the results\footnote{\small To increase the contrast of the image differences on paper, we dilate the image (with radius $2$) and adjust its pixel values to cover the range between $0$ and $1$.} of decoding an image using Python's PIL and compares it to decoding with OpenCV.
As shown, edge pixels are not decoded consistently, even though these are critical pixels for inference tasks such as object detection.

\textbf{Cropping and Resizing}\xparsep
Accuracy is sometimes reported for cropped datasets, and this is  often overlooked when evaluating a model. %
For Inception-v3, for example, input images are $87.5\%$ center-cropped and then resized to $299\times299$. 
Figure~\ref{fig:diff_output_process_pred} shows the effect of cropping on accuracy: (a) is the original image; (b) is the result of center cropping the image with $87.5\%$ and then resizing; (c) is the result of just resizing; (d) and (f) shows the top-$5$ results for images (b) and (c).
Intuitively, the effects of cropping  are more pronounced for  images  where the marginal regions are meaningful (e.g. framed paintings).

\textbf{Type Conversion and Normalization}\xparsep
After decoding, the image data is in bytes and is converted to FP32 (assuming an FP32 model). %
Mathematically, float to byte conversion is $float2byte(x) = 255x$ and byte to float conversion is $byte2float(x) = \frac{x}{255.0}$.
Because of programming semantics, however, the executed behavior of float to byte conversion is $float2byte(x) = \floor{255x}$.
The input may also be normalized to have zero mean and unit variance ($\frac{pixel-mean}{stddev}$). 
We find that the order of operations for type conversion and normalization matters.
Figure~\ref{fig:diff_output_process} shows the image processing results using different order of operations for $meanByte = stddevByte = 127.5$ and $meanFloat = stddevFloat = 0.5$ where: (a) is the original image, (b) is the result of reading the image in bytes then normalizing it with both mean and standard deviation in bytes, $byte2float(\frac{imgByte - meanByte}{stddevByte})$, (c) is the result of reading an image in floats then normalizing it with both mean and standard deviation in FP32, $\frac{byte2float(imgByte) - meanFloat}{stddevFloat}$, and (d) is the difference between (b) and (c). 
The inference results of (b) and (c) are shown in (e) and (f).

\begin{wrapfigure}{r}{0.4\textwidth}
  \vspace{-20pt}
    \setlength{\abovecaptionskip}{0pt}
  \centering%
\includegraphics[width=0.4\textwidth]{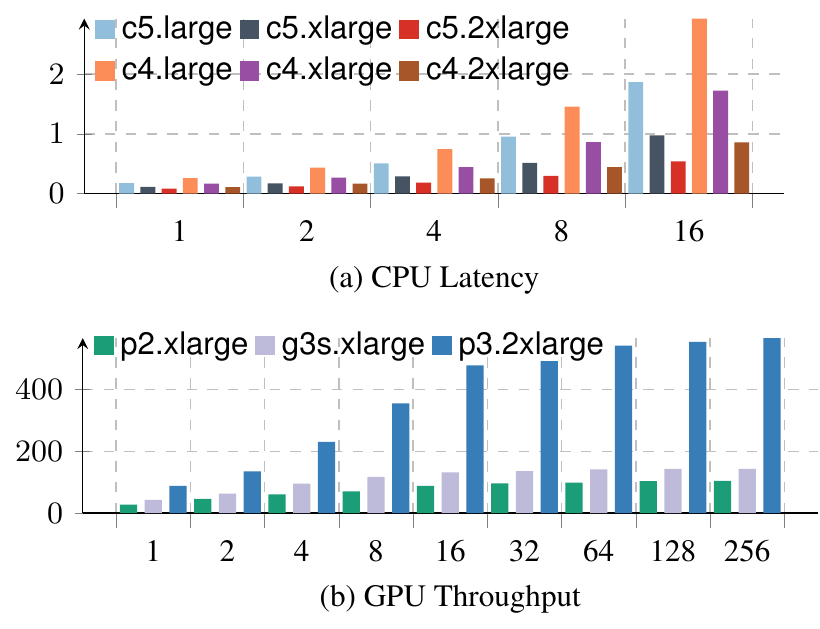}
\vspace{-10pt}
\caption{Inference latency of Inception-v3 for (a) CPU and (b) GPU systems. The $x$-axis is the batch size, and the $y$-axis is latency in seconds for  (a) and throughput in $images/second$ for (b). }
\label{fig:eval_machine}
  \vspace{-35pt}
\end{wrapfigure}

Table~\ref{tab:models}  shows the effects of pre-processing operations~\footnote{\small 
We omit from Table 1 the data layout pitfall results, since, as expected, it results in very low accuracy.} on the top 1 and top 5 accuracy for the entire ImageNet~\citep{deng2009imagenet} validation dataset.
The experiments are run in parallel on $4$ Amazon \texttt{p3.2xlarge} systems.
We can see that the accuracy errors due to incorrect pre-processing might be hard to debug, since they might only affect a small subset of the inputs.
For example, failure to center-crop the input results in $1.45\%-7.5\%$ top 1 accuracy difference, and $0.36\%-4.22\%$ top 5 accuracy difference.

\begin{table}
  \vspace{-30pt}
  \centering
    \setlength{\abovecaptionskip}{0pt}
  \resizebox{\textwidth}{!}{%
  
\begin{tabular}{lcccccccc} \toprule
\centering%

& \multicolumn{2}{c}{{{\bfseries Expected}}}
& \multicolumn{2}{c}{{{\bfseries Color Layout}}}
& \multicolumn{2}{c}{{{\bfseries Cropping}}}
& \multicolumn{2}{c}{{{\bfseries Type Conversion}}} \\

\cmidrule[0.4pt](lr{0.125em}){2-3}%
\cmidrule[0.4pt](lr{0.125em}){4-5}%
\cmidrule[0.4pt](lr{0.125em}){6-7}%
\cmidrule[0.4pt](lr{0.125em}){8-9}%

\centering%
   \textbf{Model Name} &  \textbf{Top1} & \textbf{Top5}  & \textbf{Top1} & \textbf{Top5}   & \textbf{Top1} & \textbf{Top5}   & \textbf{Top1} & \textbf{Top5}    \\ \midrule

Inception-V3~\citep{szegedy2016rethinking}  &   $78.41\%$   &  $94.07\%$   &  $67.44\%$   &  $88.44\%$  &  $78.27\%$   &  $94.24\%$   &  $78.41\%$   &  $94.08\%$  \\
MobileNet1.0~\citep{howard2017mobilenets}   &  $73.27\%$   &  $91.30\%$   &  $59.22\%$   &  $82.95\%$  &  $71.26\%$   &  $90.17\%$   &  $73.27\%$   &  $91.29\%$  \\
ResNet50-V1~\citep{he2016deep}  &   $77.38\%$   &  $93.58\%$   &  $63.21\%$   &  $85.65\%$  &  $75.87\%$   &  $92.82\%$   &  $77.40\%$   &  $93.56\%$  \\
ResNet50-V2~\citep{he2016identity}  &   $77.15\%$   &  $93.43\%$   &  $63.35\%$   &  $85.95\%$  &  $75.71\%$   &  $92.72\%$   &  $77.13\%$   &  $93.42\%$  \\
VGG16~\citep{simonyan2014very}  &  $73.23\%$   &  $91.31\%$   &  $59.17\%$   &  $82.77\%$  &  $71.71\%$   &  $90.61\%$   &  $73.24\%$   &  $91.33\%$  \\
VGG19~\citep{simonyan2014very}  &  $74.15\%$   &  $91.77\%$   &  $60.41\%$   &  $83.57\%$   &  $72.66\%$   &  $90.99\%$   &  $74.14\%$   &  $91.75\%$  \\
\bottomrule
\end{tabular}%
}
  \caption{
      The effects of the pre-processing on the Top 1 and Top 5 accuracy for heavily cited models.
  }
\setlength{\belowcaptionskip}{-5pt}
  \label{tab:models}
   \vspace{-20pt}
\end{table}

\begin{wrapfigure}{r}{0.42\textwidth}
\centering
\vspace{-40pt}
\resizebox{0.95\linewidth}{!}{
\begin{tabular}{ l  l  r r }
\centering \textbf{Instance} & \textbf{Hardware} & \textbf{\$/hr} & \textbf{Cost/Perf.} \\ \midrule
    \texttt{p2.xlarge} & Tesla K80 (Kepler), 12GB & 0.9  & 2.39 \\ 
    \texttt{g3s.xlarge} & Tesla M60 (Maxwell), 8GB & 0.75 & 1.45 \\
    \texttt{p3.2xlarge} & Tesla V100-SXM2 (Volta), 16GB & 3.06 & 1.49 \\
    \texttt{c5.large} & 2 Intel Platinum 8124M, 4GB & 0.085 & 2.76 \\
    \texttt{c5.xlarge} & 4 Intel Platinum 8124M, 8GB & 0.17  & 2.88 \\
    \texttt{c5.2xlarge} & 8 Intel Platinum 8124M, 16GB & 0.34  & 3.19 \\
    \texttt{c4.large} & 2 Intel Xeon E5-2666 v3, 3.75GB & 0.1 & 5.09 \\
    \texttt{c4.xlarge} & 4 Intel Xeon E5-2666 v3, 7.5GB & 0.199 & 5.95 \\
    \texttt{c4.xlarge} & 8 Intel Xeon E5-2666 v3, 15GB & 0.398 & 5.94  \\
	\hline
\end{tabular}
}
\vspace{-10pt}
\captionof{table}{Amazon systems used for evaluation.}
\label{tbl:systems}
  \vspace{-10pt}
\end{wrapfigure}

\subsection{Hardware Evaluation}
We use \carml{} to compare different hardware's achieved latency and throughput while fixing the model and software stack. 
We launch the same \carml{} TensorFlow agent on $9$ different Amazon EC2 systems recommended for DL (shown in Table~\ref{tbl:systems}).
These systems are equipped with either GPUs or CPUs.
We use \carml{}'s UI to run the evaluations in parallel across all $9$ systems, and measure the achieved latency and throughput of the Inception-v3 model as the batch size is varied (shown in Figure~\ref{fig:eval_machine}).
Throughput (images/second) is defined as the inverse of latency (second); i.e. throughput = $\frac{1}{\text{latency}}$.
Using the performance information in Figure 9  (maximum throughput at the optimal batch size), along with system pricing information in Table~\ref{tbl:systems}, we calculate the cost/performance as ``dollars per million images'', listed as \texttt{Cost/Perf.}in Table~\ref{tbl:systems}.
We find that for the model, software stack, and hardware used in the experimentation, GPU instances are more cost-efficient than CPU instances for batched inference.
We also observe that the \texttt{g3s.xlarge} is as cost efficient as the \texttt{p3.2xlarge}, because of the high price of the  \texttt{p3.2xlarge} instance.

\subsection{Framework Evaluation and Introspection}
We use \carml{} to compare and introspect frameworks' performance by fixing the model and hardware.
For illustration purpose, we show AlexNet, since it has less than $10$ layers and fits within the paper.
We use \carml{}'s TensorRT, MXNet, Caffe2, and Caffe agents and run them on the Amazon \texttt{p3.2xlarge} system.
Figure~\ref{fig:tensorrt_vs_caffe2_framework_compare} shows 
AlexNet's latency across frameworks.
To understand the performance of each framework, we use \carml{}'s profiler to delve deep and capture each evaluation's layer and library performance information.
Through the data, we observe that ML layers across frameworks are implemented differently and dispatched to different library functions.
Take the first \textit{conv2} and the following \textit{relu} layers for example.
In TensorRT, these two layers are fused and are mapped into two \texttt{trt\_volta\_scudnn\_128x128\_relu\_small\_nn\_v1} kernels~\citep{oyama2018accelerating} which take $1.95ms$.
In Caffe2, however, the layers are not fused and take $2.63ms$.
The sub-model profile information helps identify bottlenecks within the model inference.
We can see that \carml helps understand the performance across the HW/SW stack which is key to evaluating HW/SW stack choices.

\vspace{-3pt}

\section{Related Work}\label{sec:related}

To encourage repeatability in ML/DL research,
guidelines~\citep{mitchell2019model,showyourwork,li2019random,lipton2019research,pineau2018iclr,ReproducibilityChecklist} have been developed which authors are advised to follow.
These guidelines are checklists of what is required to ease reproducibility and encourage model authors to publish code and write down the HW/SW constraints needed to repeat the evaluation.
More often than not, model authors use notebooks~\citep{ragan2014jupyter}, package managers~\citep{ck,ck112} or containers~\citep{kurtzer2017singularity,Godlove:2019:SSS:3332186.3332192} to publish their code or specify the SW requirements.
These SW requirements are accompanied with a description of the usage, required HW stack, and are published to public repositories (e.g. on GithHub).
Through its design, \carml{} guarantees repeatable evaluations by codifying the model evaluation through the manifest and user-provided HW constraints. %

Both industry and academia have developed consortiums to build benchmark suites that evaluate widely used models~\citep{mlperf,mlmark,aimatrix,gao2019aibench,li2019acrossstack}.
These benchmark suites provide separate (non-uniform) scripts that run each model.
Each researcher then uses these scripts to perform evaluations on their target HW/SW stack.
\carml{}'s model pipeline specification overlaps with the demarcation used by other benchmark suites (e.g. \citet{mlperf} seperates model evaluation into pre-processing, model inference, and post-processing).
\carml{}, as an evaluation platform, can incorporate models from benchmark suites so that they can benefit from the distributed evaluation, profiling, and experiment management capabilities.
\carml{} currently has models from benchmark suites such as \citet{mlperf} Inference and Alibaba's \citet{aimatrix} built in.

To allow for distributed evaluation, existing platforms utilize general distributed fabrics~\citep{burns2016borg,boehm2016systemml,hindman2011mesos}  to perform model serving~\citep{kubeflow,chard2019dlhubx,novella2018container,pachyderm,zhou2019katib} or experimentation~\citep{tsay2018runway,faipep}.
\carml{} differs in that it decouples the specification and provisioning of the model evaluation pipeline from the HW/SW stack to enable repeatable and fair evaluations.
Moreover, it allows users to introspect the execution at sub-model granularity.
To the best of the author's knowledge, no previous design addresses the confluence of \feature{1-3} repeatability, fairness, and introspection within scalable model evaluation at the same time.

\vspace{-3pt}

\section{Conclusion}\label{sec:conclusion}

Everyday, an increasingly complex and diverse DL models as well as hardware/software (HW/SW) solutions are proposed --- be it algorithms, frameworks, libraries, compilers, or hardware.
Both industry and research are hard-pressed to quickly, thoroughly, consistently, and fairly evaluate these new innovations.
This paper proposes \carml, which is a specification along with a distributed runtime design that is scalable, extensible, and easy-to-use.
Through \carml, users can perform fair and repeatable comparisons across models, software stacks, and hardware.
\carml's careful design of the specification, runtime, and parallel evaluation flow reduces time-to-test for model evaluators.
With \carml, we evaluate a set of representative image classification models and present insights into how different pre-processing operations, hardware, and framework selection affect model accuracy and performance.

\bibliography{main}
\bibliographystyle{iclr2019_conference}

\clearpage
\newpage

\appendix
\section{Supplementary Material}

\carml{} is a big system, and we had to selectively choose topics due to the space limitation.
This supplementary materials section is used to provide details about \carml{} that we were unable to cover in the paper's main body.
Specifically, we discuss how \carml:
(a) incorporates the latest research and production models through model manifests by showing object detection and instance segmentation models.
 (b) Attracts users by providing a web interface and command line for scalable model evaluation. 
(c) Is built from a set of modular components which allows it to be easily customized and extended.

\subsection{\carml{} Model Manifests}

Listing~\ref{lst:ssd-mobilenet-coco} shows the manifest of \texttt{SSD\_MobileNet\_v1\_COCO}, an objection detection model, for TensorFlow.
This model embeds the pre-processing operations in the model graph, and thus requires no normalization, cropping, or resizing.
The major difference from a image classification model manifest is the task type (being \texttt{object\_detection}) and the outputs.
There are three output tensors for this model (boxes, probabilities, and classes).
These output tensors are processed by \carml{} to produce a single object detection feature array, which can then be visualized or used to calculate the metrics (e.g. mean average precision).
 
\begin{lstlisting}[
    language=yaml,
    escapeinside={(*}{*)},
    floatplacement=H,
    captionpos=b,
    label=lst:ssd-mobilenet-coco,
    caption={\carml's model specification for \texttt{SSD\_MobileNet\_v1\_COCO} TensorFlow model.
    },
    escapechar=|
]
name: SSD_MobileNet_v1_COCO  # name of your model
version: 1.0 # version information in semantic version format
task: object_detection # task type
framework:
  name: TensorFlow # framework name
  version: 1.12.x # framework version contraint
container: # containers used to perform model evaluation
  amd64:
    gpu: mlcn/tensorflow:amd64-cpu
    cpu: mlcn/tensorflow:amd64-gpu
  ppc64le:
    cpu: mlcn/tensorflow:ppc64le-gpu
    gpu: mlcn/tensorflow:ppc64le-gpu
description: ...
references: # references to papers / websites / etc.. describing the model
  - ...
license: Apache License, Version 2.0 # license of the model
inputs: # model inputs
  - type: image # first input modality
    element_type: uint8
    layer_name: image_tensor
    layout: HWC
    color_layout: RGB
outputs:
  - type: box
    element_type: float32
    layer_name: detection_boxes
  - type: probability
    element_type: float32
    layer_name: detection_scores
  - type: class
    element_type: float32
    layer_name: detection_classes
    features_url: https://.../labels.txt
source:
  graph_path: https://.../ssd_mobilenet_v1_coco_2018_01_28.pb
attributes: # extra model attributes
  training_dataset: COCO # dataset used to for training
  manifest_author: ...
\end{lstlisting}

Listing~\ref{lst:ssd-mask-rcnn-coco} shows the manifest of \texttt{Mask\_RCNN\_ResNet50\_v2\_Atrous\_COCO}, an instance segmentation model, for MXNet.
The major difference from the object detection model in Listing~\ref{lst:ssd-mobilenet-coco} is the task type (being \texttt{instance\_segmentation}) and the outputs.
Listing~\ref{lst:ssd-mask-rcnn-coco} shows four outputs for this model (boxes, probabilities, classes, and masks).
These output tensors are  processed by \carml{} to produce a single instance segmentation feature array.
Note that unlike TensorFlow, MXNet uses layer indices in place of layer names to get the tensor objects.

\begin{lstlisting}[
    language=yaml,
    escapeinside={(*}{*)},
    floatplacement=H,
    captionpos=b,
    label=lst:ssd-mask-rcnn-coco,
    caption={\carml's model specification for  \texttt{Mask\_RCNN\_ResNet50\_v2\_Atrous\_COCO} MXNet model.
    },
    escapechar=|
]
name: Mask_RCNN_ResNet50_v2_Atrous_COCO # name of your model
version: 1.0 # version information in semantic version format
task: instance_segmentation
framework:
  name: MXNet # framework for the model
  version: 1.4.x # framework version contraint
container: # containers used to perform model evaluation
  amd64:
    gpu: mlcn/mxnet:amd64-cpu
    cpu: mlcn/mxnet:amd64-gpu
  ppc64le:
    cpu: mlcn/mxnet:ppc64le-gpu
    gpu: mlcn/mxnet:ppc64le-gpu
description: ...
references: # references to papers / websites / etc.. describing the model
  - ...
license: Apache License, Version 2.0 # license of the model
inputs: # model inputs
  - type: image # first input modality
    element_type: uint8
    layout: HWC
    color_layout: RGB
outputs:
  - type: box
    element_type: float32
    layer_name: 0
  - type: probability
    element_type: float32
    layer_name: 1
  - type: class
    element_type: float32
    layer_name: 2
    features_url: https://.../labels.txt
  - type: mask
    element_type: float32
source: # specifies model graph and weights sources
  base_url: http://.../mxnet/Mask_RCNN_ResNet50_v2_Atrous_COCO/
  graph_path: model-symbol.json
  weights_path: model-0000.params
attributes: # extra model attributes
  training_dataset: COCO # dataset used to for training
  manifest_author: ...
\end{lstlisting}

\begin{figure}[h] 
    \centering
    \includegraphics[width=\textwidth]{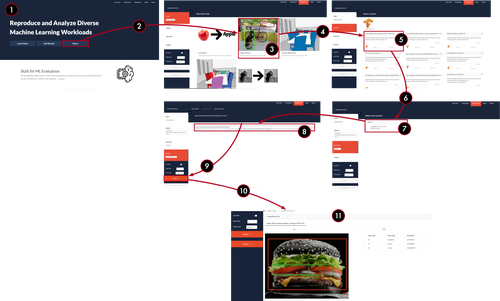}
    \caption{The \carml{} website provides an intuitive interface to conduct experiments.}
    \label{fig:website_flow}
    \vspace{-10pt}
\end{figure}

\subsection{Website Workflow}

Although \carml{} provides both command line and library interfaces, we find the website provides an intuitive flow for specifying and running experiments.
Figure~\ref{fig:website_flow} shows the flow, and a video demonstrating it can be found at \url{https://drive.google.com/open?id=1LOXZ7hs_cy-i0-DVU-5FfHwdCd-1c53z}.
In figure~\ref{fig:website_flow}, users first arrive at  \circled{1} \carml's landing page.
The landing page contains a description of the project along with links to how to setup and install \carml.
Users can try \carml{} by \circled{2} clicking the demo button, which then displays \circled{3} the inference tasks exposed through the website.
If a user \circled{4} selects object detection, then \circled{5} models that are available for object detection are displayed.
A user can then \circled{7}  selects one or more models and \circled{8} selects one or more systems to run the evaluation on.
The input can be specified as a URL, data from disk, or dataset \circled{8} and once complete the user can perform the evaluation \circled{9}.
This \circled{10} will run the evaluation on the remote system and \circled{11} display the evaluation results along with summary of the execution flow.

\begin{wrapfigure}{tr}{0.4\textwidth}
  \centering%
    \setlength{\abovecaptionskip}{0pt}
  \includegraphics[width=0.38\textwidth]{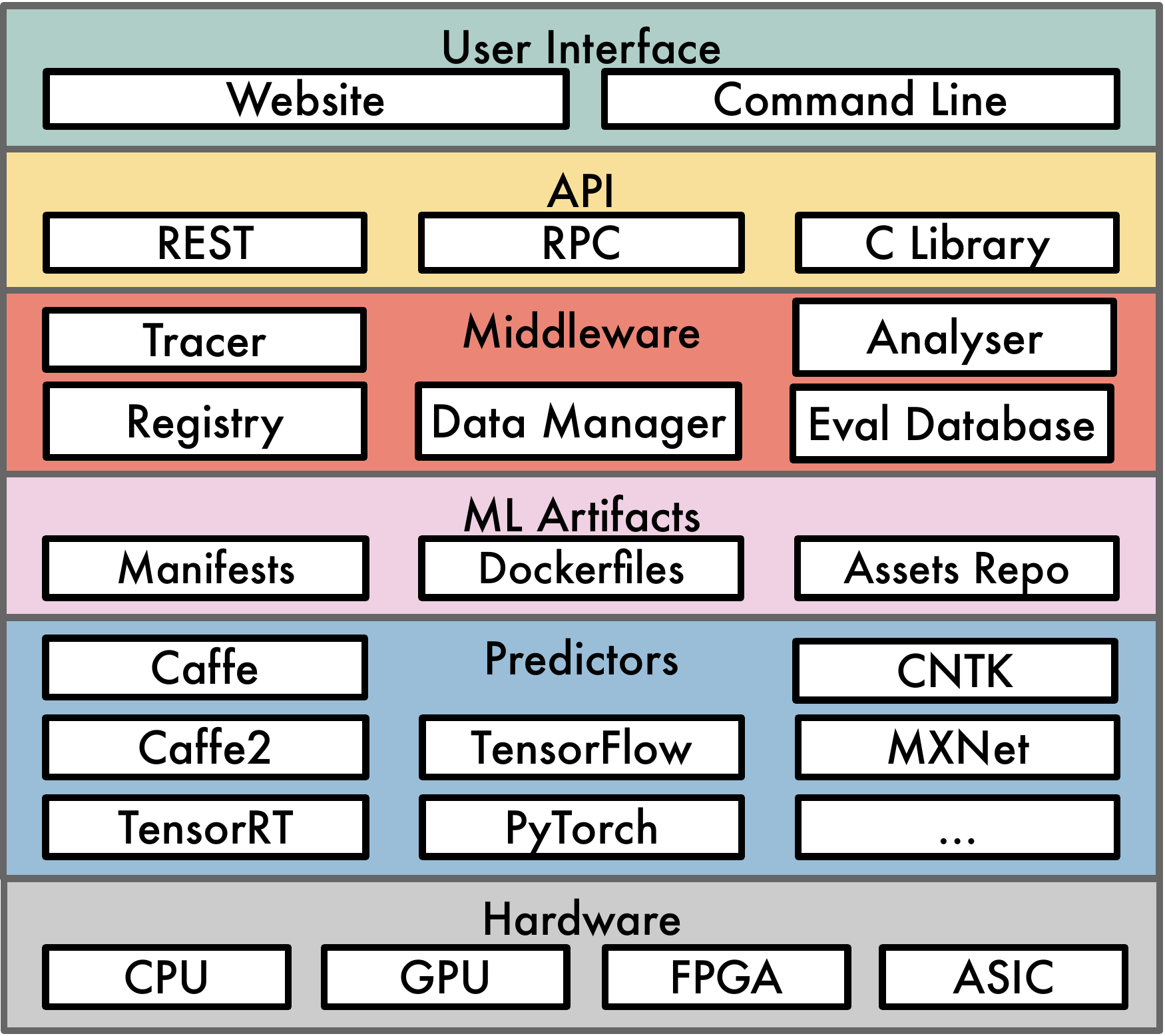}
  \caption{\carml's runtime components.}
  \label{fig:arch2}
 \vspace{-30pt}
\end{wrapfigure}

\subsection{\carml{}'s Runtime Architecture}

In this section we describe each component in Figure~\ref{fig:arch2} in detail.
The runtime is designed to be extensible and customizable.

\subsubsection{User Interface and API}

\carml can be used as an application or as a library.
Users interact with \carml application through its website, command line, or its API interface.
The website and command line interface allow users to evaluate and profile models without familiarity with the underlying frameworks or profiling tools.
Users who wish to integrate \carml within their existing tools or pipelines can use the REST or RPC APIs.
They can also compile \carml as a standalone shared library and use it within their C/C++, Python, or Java projects.

\subsubsection{ML Artifacts}
As discussed in the main body of the paper, replication of model accuracy and performance results is dependent on: the usage of specific HW/SW stack; the training dataset; and the pre/post-processing steps on the inputs and outputs.
\carml specifies these requirements via a model manifest file described in Section~\ref{sec:design}.
The manifest tells \carml the HW/SW stack to instantiate and how to evaluate the model.

\textbf{Asset Versioning} --- Models, frameworks, and datasets are versioned using a semantic versioning scheme. 
The \carml middleware layer uses this information for asset management and discovery.
To request a model, for example, users specify model, framework, hardware, or dataset constraints. %
\carml solves the constraint and returns the predictors (systems where the model is deployed) that satisfy the constraint.
The model evaluation can then be run on one of (or, at the user request, all) the predictors.

\textbf{Docker Containers} --- To maintain the SW stack, evaluations are performed within docker containers.
To facilitate user introspection of the SW stack, \carml integrates with existing docker tools that allows querying images's SW environment and metadata.

\textbf{Pre/Post-Processing Operations} --- \carml provides the ability to perform common operations such as resizing, normalization, and scaling without writing code.
It also allows users to specify code snippets for pre/post-processing within the manifest file which are run within a Python subsession.
\carml is able to support a wide variety of models for different input modalities.

\textbf{Evaluation History} --- \carml uses the manifest information as keys to store the evaluation results in a database.
Users can view historical evaluations through the website or command line using query constraints similar to the ones mentioned above.
\carml summarizes and generates plots to aid in comparing the performance across experiment.

\subsubsection{Framework and Model Predictors}

A predictor is a thin abstraction layer that exposes a framework through a common API.
A predictor is responsible for evaluating models (using the manifest file) and capturing the results along with the framework's profile information.%
A predictor publishes its HW/SW stack information to \carml's registry at startup, can have multiple instantiations across the system, and is managed by \carml's middleware.

\subsubsection{Middleware}

The middleware layer is composed of services and utilities for orchestrating, provisioning, aggregating, and monitoring the execution of predictors --- acting  as a conduit between the user-facing APIs and the internals of the system.

\textbf{Manifest and Predictor Registry} ---
\carml uses distributed key-value database to store the registered model manifests and running predictors.
 \carml leverages the registry to facilitate discovery of models, load balancing request across predictors, and to solve user constraint for selecting the predictor (using HW/SW stack information registered).
The registry is dynamic --- both model manifests and predictors can be added or deleted at runtime throughout the lifetime of the application.

\textbf{Data Manager} ---
\carml data manager is responsible for downloading the assets (dataset and models) required by the model's manifest file.
Assets can be hosted within \carml's assets repository, or hosted externally.
For example, in Listing~\ref{lst:model_manifest}  ( Lines~\ref{line:start_weights}--\ref{line:end_weights}) the manifest uses a model that's stored within the \carml repository, the data manager downloads this model on demand during evaluation.

Within \carml's repository, datasets are stored in an efficient data format and are placed near compute on demand.
The dataset manager exposes a consistent API to get values and iterate through the dataset.

\textbf{Tracer} --- 
The \carml tracer is middleware that captures the stages of the model evaluation, leverages the predictor's framework profiling capability, and interacts with hardware and system level profiling libraries to capture fine grained metrics.
The profiles do no need to reflect the wall clock time, for example, users may integrate a system simulator and publish the simulated time rather than wall-clock time.

\carml publishes the tracing results asynchronously to a distributed server --- allowing users to view a single end-to-end time line containing the pipeline traces.
Users can view the entire end-to-end time line and can  ``zoom'' into specific component (shown in Figure~\ref{fig:profile_levels}) and traverse the profile at different abstraction levels.
To reduce trace overhead, users control the granularity (AI component, framework, library, or hardware) of the traces captured.

\carml leverages off-the-shelf tools to enable whole AI pipeline tracing.
To enable the AI pipeline tracing, users inject a reference to their tracer as part the  model inference API request to \carml.
\carml then propagates its profiles to the injected application tracer instead of the \carml tracer --- placing them within the application time line.
This allows \carml to integrate with existing application time lines and allows traces to span API requests.

\subsubsection{Frameworks}

At time of writing, \carml has built-in support for
Caffe, Caffe2, CNTK, MXNet, PyTorch, Tensorflow, TFLite, and TensorRT.
\carml uses ``vanilla'' unmodified versions of the frameworks and uses facilities within the framework to enable layer-level profiling --- this allows \carml to work with binary versions of the frameworks (version distributed through Python's pip, for example) and support customized or different versions of the framework with no code modifications.
To avoid overhead introduced by scripting languages, \carml's supported frameworks use the frameworks' C-level API directly --- consequently the evaluation profile is as close to the hardware as possible.

\subsubsection{Hardware}

\carml has been tested on X86, PowerPC, and ARM CPUs as well as NVIDIA's Kepler, Maxwell, Pascal, and Volta GPUs.
It can leverage NVIDIA Volta's TensorCores, and can also perform inference with models deployed on FPGAs.
During evaluation, users can select hardware constraints such as: whether to run on CPU or GPU, type of architecture, type of interconnect, and minimum memory requirements --- which \carml considers when selecting a system.

\subsection{\carml{} Source Code}

This project is open-source and the code spans multiple ($>15$) repositories on GitHub.
Thus it is difficult to anonymize for the blind review process.
We are happy to share the links to the source code with the PC members. The links will be included in the paper after the blind review process.

\end{document}